\documentstyle[psfig,aprm2002]{article}
\title{%
The Central Parsecs of the Bright Quasar PKS~1921--293}
\authors{% DO NOT DELETE THIS LINE.
Z.-Q. Shen\affilmark{1},
J. M. Moran\affilmark{2},
K. I. Kellermann\affilmark{3}
}% DO NOT DELETE THIS LINE.

\affiltexts{% DO NOT DELETE THIS LINE.
\affiltext{1}{Institute of Space and Astronautical Science, Yoshinodai
3-1-1, Sagamihara, Kanagawa 229-8510, JAPAN},
\affiltext{2}{Harvard-Smithsonian Center for Astrophysics, 60 Garden Street, Cambridge, MA 02138, USA},
\affiltext{3}{National Radio Astronomy Observatory, 520 Edgemont Road, Charlottesville, 
VA 22903, USA},
}% DO NOT DELETE THIS LINE.

\firstauthor
{ZS}
{zshen@vsop.isas.ac.jp}

\papernoADS{%(for example PL 1-1)
PL1-1
}% DO NOT DELETE THIS LINE.
\authorsADS{%
Shen, Z.-Q.;
Moran, J. M.;
Kellermann, K. I.;
}
\affiliationsADS{%
AA(ISAS)
AB(CfA)
AC(NRAO)
}

\begin{document}

%%%%%%%%%%%%%%%%%%%%%%%%%%%%%%%%%%%%%
% Abstract
%
\begin{abstract}
\vspace*{0.25cm}
We report on a VLBA imaging study of the nearby bright southern blazar 
PKS~1921--293 (OV--236). 
High resolution VLBA observations, made at four frequencies 
(5, 12, 15, and 43~GHz) over the period 1994-2000, have  
revealed a strongly curved jet extending out to about 50 parsecs 
from the presumed central engine. 
Two epoch VLBA observations, each simultaneously carried out at both 5 
and 43\,GHz, show a large position angle difference of 
51$^\circ$ -- 67$^\circ$ between the jet emission at 5 and 43\,GHz.
Although the core of PKS~1921--293 has one of the highest brightness 
temperatures measured in any compact radio source, unlike other 
bright blazars it is not a source of $\gamma$--ray emission. 
However, there is evidence in these images for superluminal motion 
within the central region (a few parsecs from the core) and within 
the north-east diffuse emission region.
In all six-epoch 43\,GHz images, two equally compact bright components
within the central parsec are seen. \\

\end{abstract}

%%%%%%%%%%%%%%%%%%%%%%%%%%%%%%%%%%%%%
% General
%

\section{Introduction} 
\vspace*{0.25cm}
PKS~1921--293 (OV--236) is identified with a 17.5 V-magnitude quasar.
At a redshift of 0.352 (Wills \& Wills 1981), PKS\,1921$-$293 is one of the closest 
members of its class.
An angular resolution of 1~mas corresponds to a linear resolution of 4.6~pc 
(assuming H$_0$~=~65~km~s$^{-1}$~Mpc$^{-1}$ and q$_0$~=~0.5).

PKS~1921--293 is one of the strongest and most compact extragalactic radio sources known,
which makes it a prime candidate for high-resolution VLBI observations.
PKS~1921--293, together with 3C\,273B and 3C\,279, is currently among the brightest 
extragalactic sources 
in the sky at millimeter wavelengths, having a flux density at 3\,mm greater than 
5\,Jy since 1990 with a peak of 15.3\,Jy in April 1994 (Tornikoski et al. 1996). Its compactness, 
implied by its flat radio spectrum, has been confirmed by space VLBI (VSOP) 
observations (Shen et al. 1999). On projected space-ground baselines of 25,000\,km (about 
three times longer than the longest ground baselines), PKS~1921--293 had a correlated flux 
density of 1.0\,Jy (1.6\,GHz), and a derived core brightness temperature limit 
of 3.0$\times10^{12}$\,K (in the rest frame of the quasar). \\

%%%%%%
%%%%%%
% Figure 1
%
\begin{figure}[ht]
\begin{center}
\vspace*{-1cm}
\leavevmode\psfig{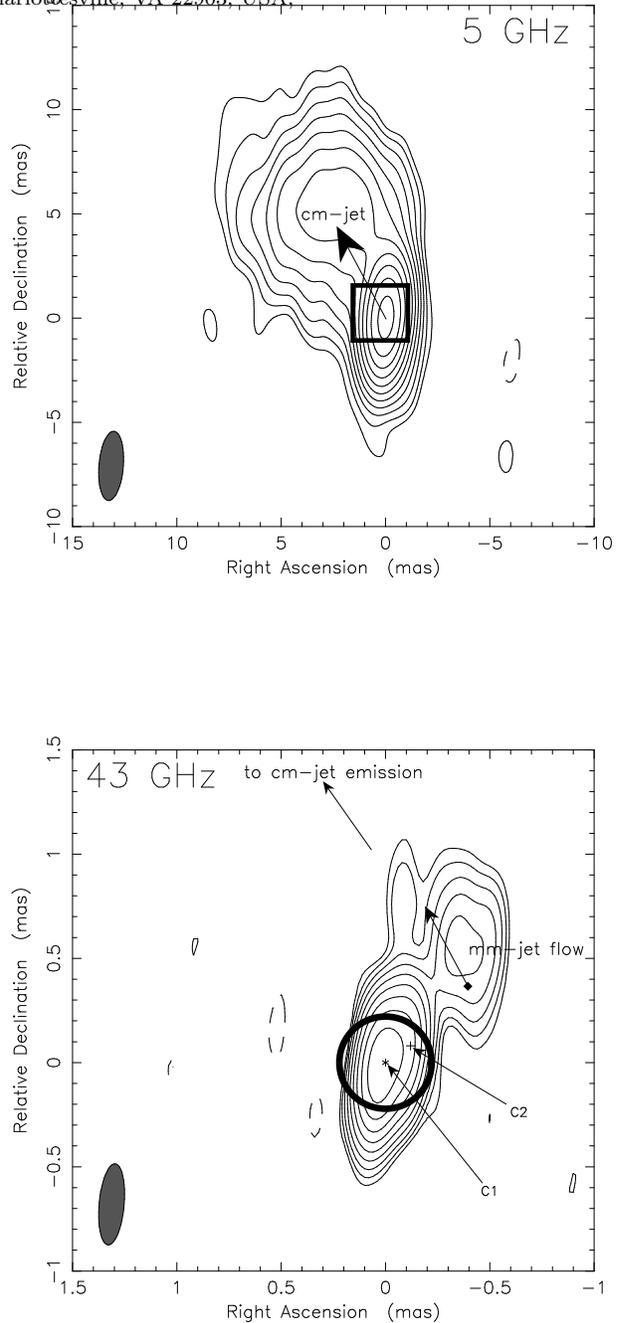}
\end{center}
\caption{Two VLBA images of PKS\,1921--293 simultaneously observed at 5\,GHz (upper panel) and
43\,GHz (lower panel) in September 2000. 
Note the difference in the scale since the 43\,GHz image occupies only
1\% area (within the {\bf bold} square surrounding the origin) of the 5\,GHz image.
The part of the mm-jet flow from which the proper motion was obtained is indicated by a vector 
with the starting position (filled diamond in 43\,GHz image) detected at epoch 1994.32 
(see section 4). This mm-jet component is not seen in the 2000 image.
For a discussion of components C1 and C2 within the central region of radius 1\,pc  
({\bf bold} circle in 43\,GHz image), see section 5.}
\label{fig:1}
\end{figure}
%%%%%%
%%%%%%

\section{Observations and Imaging}
\vspace*{0.25cm}
The new observations of PKS\,1921--293 were made with the NRAO\footnote{The National Radio 
Astronomy Observatory (NRAO) is operated by Associated Universities, Inc., under cooperative 
agreement with the National Science Foundation.} VLBA at four frequencies 
(5, 12, 15, and 43\,GHz) over a period of six years from 1994 to 2000. The data were
correlated at the VLBA correlator in Socorro, New Mexico, USA. Standard VLBI data calibrations 
and corrections for residual delays and rates were done in AIPS. At 43\,GHz, the 
atmospheric opacity correction was also performed in addition to a priori amplitude 
calibration. The final image processing was carried out using AIPS and DIFMAP.

In general, the source morphologies obtained at various centimeter wavelengths (i.e., 5, 12 
and 15\,GHz) are very similar and consistent with the past observations. However, their overall 
structural orientation shows a significant difference from that observed at 43\,GHz.
Figure\,1 shows two images
observed simultaneously at 5 and 43\,GHz in September 2000. At centimeter wavelengths 
all of the VLBA images are characterized by a diffuse emission region 
(cm-jet) extended about 10\,mas along a position angle (P.A.) of 30$^\circ$ as well as a very compact, 
bright core at the center. At 43\,GHz, our six-epoch VLBA images reveal an even more compact core 
emission region plus the jet emission (mm-jet). The mm-jet has a varying P.A. with time, but 
is oriented toward the north-west. The core brightness temperature derived from 43\,GHz
images has a lower limit between 10$^{11}$ to 10$^{12}$\,K.

\clearpage

\section{Curvature}
\vspace*{0.01cm}
One of the important results of the analysis of the multi-epoch, multi-frequency VLBA images of
PKS\,1921--293 is its strongly 
curved jet emission. The difference in P.A. between
the cm-jet and the mm-jet is at least 51$^\circ$\,--\,67$^\circ$, as determined by the
simultaneous two-frequency (5 and 43\,GHz) observations made both in February 1996 
and in September 2000 (see Figure\,1). 
Such a large bending is not unexpected for a jet that is beamed close to the observer's line of 
sight, for which small changes in angle can be amplified by projection effects. 

The 43\,GHz VLBA observations of the mm-jet and its motions (see section 4 below) suggest a 
curved trajectory within the central few parsecs, while the trajectory of the cm-jet appears 
to be constant at larger core separations 
(seen at lower frequencies). Figure\,2 is a schematic diagram showing a single smoothly bent path 
connecting the mm-jet and the cm-jet through an inner jet component detected by the VSOP 
observations (Shen et al. 1999). 

It should be emphasized that whether the curvature in the jet emission is responsible 
for the non-detection of the $\gamma$--ray emission by the EGRET (Fichtel et al. 1994) is not clear,
as some larger bending angles have also been seen in other $\gamma$--ray bright sources 
(Britzen et al. 2001). 

%%%%%%
%%%%%%
% Figure 2
%
\begin{figure}[ht]
\vspace*{-1.35cm}
\begin{center}
\leavevmode\psfig{file=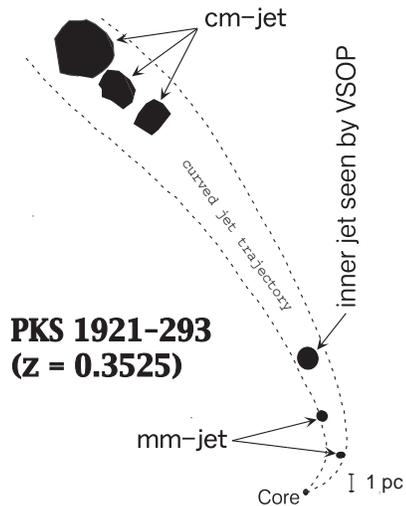,height=80mm,width=80mm}
\end{center}
\vspace*{-0.5cm}
\caption{A schematic model showing how the PKS\,1921--293 jet emission observed by the
VLBA at centimeter- and millimeter-wavelengths and by VSOP at 1.6\,GHz are connected
via a single smoothly curved trajectory.}
\label{fig:2}
\end{figure}
%%%%%%
%%%%%%

\vspace*{-0.85cm}
\section{Motion}

For the large scale cm-jet emission, we obtained the proper motion from the position of the 
brightest component at 5\,GHz that has a mean P.A. of 28$^\circ$$\pm$2$^\circ$.
The fitted proper motion over the past 7.5 years is 0.15$\pm$0.08\,mas\,yr$^{-1}$,
corresponding to a superluminal velocity of 3.0$\pm$1.6\,c.
The proper motion (apparent superluminal velocity) within the mm-jet is estimated to be 
0.16$\pm$0.01\,mas\,yr$^{\tiny -1}$ (3.2$\pm$0.2\,c), 
along a position angle of 28$^\circ$$\pm$3$^\circ$ with regard to a reference
position at epoch 1994.32 of 0.54\,mas and P.A.\,=\,$-47^\circ$ (from the origin).
The P.A. of the cm-jet emission region relative to this reference mm-jet component 
is about 34$^\circ$$\pm$3$^\circ$. Note that the error bars are derived from the statistical 
uncertainties only. Therefore, the jet components in PKS\,1921--293 seem to move
along a straight path from about 2.5\,pc to about 25\,pc (as depicted in Figure\,2). 
The P.A. of the reference mm-jet emission is roughly perpendicular to the derived direction
of its superluminal motion.
This may further suggest that the jet emission observed so far is part of a curved trajectory 
that undergoes a sharply bending ($\Delta$PA\,$\approx$\,90$^\circ$) phase within the central 2\,pc region. 

\section{The Innermost ($<$ 1\,pc) Region}
\vspace*{0.25cm}
An important question is how the jet moves in the innermost region
of the central engine where the relativistic plasma is accelerated and collimated.
Our six epoch 7\,mm VLBA observations found consistently
 that the central parsec region ($\la$ 1\,pc)
consists of two equally compact components (labelled as
C1 and C2 in Figure\,1). The fitted sizes for both C1 and C2 are 0.1\,mas or less.
Away from the central parsec region, the jet emission gradually decreased in 
flux density as it was moving away.
The scenario in the central 1\,pc region, however, is quite different.
The flux density of component C1 appears not to change, compared to the
rapid variation in the total flux densities observed at 90 and 230\,GHz
by SEST (Tornikoski 2001, private communication).
On the other hand, component C2 undergoes a significant ($\sim$200\%) variation in  
flux density which seems to correlate with the
total flux density variation.

The dramatic variability seen in component C2 may argue that C2 is related
to the central activity, though the observed superluminal motion suggests 
that the component C1, at the eastern end, is the location of the core. 

Over about 6.5 yrs of six epoch 43\,GHz VLBA observations, the relative 
position between C1 and C2 of 0.2\,mas (i.e., $\la$ 1\,pc) along a P.A. of $-$50$^\circ$
remains unchanged within an uncertainty of 0.05\,mas.
High resolution VSOP observations at 5\,GHz have also suggested the existence of
such a compact double structure, but with a larger separation of
$\sim$0.4\,mas at a similar P.A. (Shen et al. 2000). 
The difference in the separation may be due to the opacity effects, 
i.e., the frequency-dependent shift of the
synchrotron self-absorbed core at longer wavelength. 
PKS\,1921--293 becomes optically thick at centimeter wavelengths

Some possible explanations are: 1) C2 is just a stationary knot that is located
at the turning point of the bent path and its emission is amplified by relativistic 
beaming; 2) C2 is a trailing shock following 
the steady, fast superluminal motion at the 
front as seen in the radio galaxy 3C\,120 (G\'{o}mez et al. 2001); 3) this may infer the 
existence of a binary system within the central parsec region of PKS\,1921--293, which 
could be associated with a massive black hole binary. 
A supermassive binary in an AGN will result in the precession
of the jet ejection axis, which may explain the observed large
bend (Begelman, Blandford \& Rees 1980).
New high-resolution VLBI imaging observations at both 86 and 43\,GHz are underway to 
investigate the binary hypothesis as well as the above-mentioned possibilities. \\

%\vspace{-1mm}
%%%%%%%%%%%%%%%%%%%%%%%%%%%%%%%%%%%%%
% References
%
\section*{References} 
\vspace*{0.25cm}

\reference
Begelman, M. C., Blandford, R. D. \& Rees, M. J. 1980, Nature, 287, 307

\reference
Britzen, S., Vermeulen, R. C., Taylor, G. B., Campbell, R. M., Browne, I. W., Wilkinson, P.,
Pearson, T. J. \& Readhead, A. C. S. 2001, in Galaxies and Their Constituents at the Highest 
Angular Resolution (IAU Symp. 205), ASP Conf. Series Vol. 205, eds. R. T. Schilizzi, S. Vogel, 
F. Paresce \& W. Elvis, (San Francisco: ASP), 106

\reference
Fichtel, C. E., Bertsch, D. L., Chiang, J., Dingus, B. L.,
 Esposito, J. A., Fierro, J. M., Hartman, R. C., Hunter, S. D.,
 Kanbach, G., Kniffen, D. A., Kwok, P. W., Lin, Y. C., Mattox, 
 J. R., Mayer-Hasselwander, H. A., McDonald, L., Michelson,
  P. F., von Montigny, C., Nolan, P. L., Pinkau, K., Radecke, 
 H.-D., Rothermel, H., Sreekumar, P., Sommer, M., Schneid, 
 E. J., Thompson, D. J. \& Willis, T. 1994, ApJS, 94, 551
 
\reference
G\'{o}mez, J.-L., Marscher, A. P., Alberdi, A., Jorstad, S. J. \& Agud\'{o}, I.
2001, ApJL, 561, L161

\reference
Shen, Z.-Q., Edwards, P. G., Lovell, J. E. J., \& Kameno, S. 2000, in Astrophysical
Phenomena Revealed by Space VLBI, eds. H. Hirabayashi, P. G. Edwards \& D. W. Murphy,
(Sagamihara, ISAS), 155.

\reference
Shen, Z.-Q., Edwards, P. G., Lovell, J. E. J., Fujisawa, K., Kameno, S. \& Inoue, M. 1999
PASJ, 51, 513

\reference
Tornikoski, M., Valtaoja, E., Ter\"{a}sranta, H., Karlamaa, K., Lainela, M., Nilsson, K., 
Kotilainen, J., Laine, S., L\"{a}hteenm\"{a}ki, A., Knee, L. B. G. \& Botti, L. C. L. 1996,
A\&AS, 116, 157

\reference
Wills, D. \& Wills. B. J. 1981, Nature, 289, 384

\vspace{-1mm} % TO MAKE THE END OF THE ARTICLE FLAT

\end{document}